# Characterizing Solar Spicules and their Role in Solar Wind Production using Machine Learning and the Hough Transform

R. Sadeghi[1] & E. Tavabi[1]

[1]Physics Department, Payame Noor University, Tehran, Iran, 19395-3697

**Abstract.** Solar winds originate from the Sun and can be classified as fast or slow. Fast solar winds come from coronal holes at the solar poles, while slow solar winds may originate from the equatorial region or streamers. Spicules are jet-like structures observed in the Sun's chromosphere and transition region. Some spicules exhibit rotating motion, potentially indicating vorticity and Alfven waves. Machine learning and the Hough algorithm were used to analyze over 3000 frames of the Sun, identifying spicules and their characteristics. The study found that rotating spicules, accounting for 21% at the poles and 4% at the equator, play a role in energy transfer to the upper solar atmosphere. The observations suggest connections between spicules, mini-loops, magnetic reconnection, and the acceleration of fast solar winds. Understanding these small-scale structures is crucial for comprehending the origin and heating of the fast solar wind.

## 1. Introduction

Spicules are spike-like structures found in the Sun's chromosphere and transition region. They exhibit spiral motion and opposite blue/red shifts. Some spicules are associated with vorticity movements and may contribute to the propagation of Alfven rotational waves. High-resolution images reveal rotating spicules, while lower-resolution observations show evidence of twisted spicules in spectral data. Solar wind generation, especially fast solar winds, is linked to coronal holes at the solar poles. Spicules appear as single jets that split into multiple formations and exhibit spinning and rotating motions. Spectroscopic studies confirm rotation and twisting in spicules, indicating complex dynamics and the presence of torsional Alfvén waves. Twisted spicules resembling solar tornadoes have been directly detected. The twist in spicules may be transferred from erupting twisted magnetic fields to the surrounding open magnetic field through external reconnection, suggesting a connection between magnetic field dynamics and spicule behavior (Pike and Mason 1998; Moore et al. 1977; Tavabi et al. 2010, 2012b; Tavabi 2014; Tavabi et al. 2014; Ajabshirizadeh et al. 2008).

## 2. Observation

This study combines data from the IRIS and Hinode/SOT instruments to investigate the dynamics of solar jets and rotational spicules. The IRIS data provide high-resolution and wide-spectral-range observations, while the Hinode/SOT data offer insights into the formation and twisting of spicules. The data are calibrated using standard software for accuracy.

## 3. Method

This study employed data from the IRIS and Hinode/SOT instruments to investigate solar jets and rotational spicules. A combination of machine learning and the Hough transform algorithm was used to separate and analyze the spicules observed in the spectra (figure 1) . The data covered various regions on the Sun, and the IRIS data had high spatial resolution and wide spectral coverage. The Hough transform algorithm identified spicule line segments in the spectra, while Doppler maps revealed the spicules' Doppler shifts(figure 2). Machine learning models were trained to classify spicules as twisted or non-twisted based on spectral, spatial, and temporal characteristics. The accuracy of the models was evaluated using test data, and the best-performing models were used to identify twisted spicules in the analysis data (figure 1). The study provided insights into the dynamics and properties of solar jets and rotational spicules through a combination of observational data, image processing techniques, and machine learning algorithms (Sadeghi and Tavabi 2022b,a; Tavabi and Koutchmy 2012; Tavabi et al. 2012a).

## 4. Results

In this study, researchers analyzed a dataset of approximately 11,500 spectral images to identify and categorize spicules using trained models and the Hough transform algorithm. They found that twisted spicules accounted for around 21% of spicules in the polar regions and 4% in the equatorial region. There were slightly more twisted spicules on the west side of the Sun compared to the east side, and the number of twisted spicules was similar at the north and south poles. Tables 1, and 2 in the study provided information on the average rotational velocity, angular deviation, and Doppler velocities of the detected twisted spicules. The num- ber of twists observed in spicules varied depending on their diameter, ranging from at least one turn for fragile structures to broader structures. The study suggested that the expansion of chromospheric mini-loops could lead to micro-eruptions similar to spicule eruptions of the spine type, based on the conservation of rotation or helicities. The structure of the transition region and low corona was found to be influenced by the complex magnetic field configuration, with higher-order multipole components rather than a simple magnetic dipole configuration. The presence of multiple magnetic mini-loop systems anchored at the photosphere resulted in a series of changes in magnetic polarity. Overall, this study provided insights into the preva- lence and characteristics of twisted spicules in different regions of the Sun and highlighted the complex magnetic field configuration governing the solar atmosphere (figure 3).

**Table 1.** The summary of data received from machine learning model analysis varies depending on the study region

|   | spicules | West Equator | East Equator | North Pole | South Pole |
|---|---|---|---|---|---|
| Number of | identified | 1843 | 1152 | 2576 | 6608 |
|   | twisted | 79 | 45 | 519 | 1434 |
|   | ratio(%) | 4.30 | 3.91 | 20.15 | 21.71 |
| Red rotational velocity of (km/s) | twisted | 25 | 24 | 36 | 34 |
| Blue rotational velocity of (km/s) | twisted | 29 | 27 | 36 | 33 |
| rotation angle of (degree) | identified | 8.0 | 7.7 | 12.3 | 12.0 |
|   | twisted | 7.3 | 7.2 | 12.3 | 12.3 |

**Table 2.** The classified data collected via machine learning model analysis.

|   | twisted spicules | | |
|---|---|---|---|
|   | average | min | max |
| poles | | | |
| rotational velocity (km/s) | 36 | 24 | 65 |
| rotation angle (degree) | 12.3 | 3.5 | 15.0 |
| equator | | | |
| rotational velocity (km/s) | 27 | 24 | 53 |
| rotation angle (degree) | 7.3 | 2.1 | 11.5 |

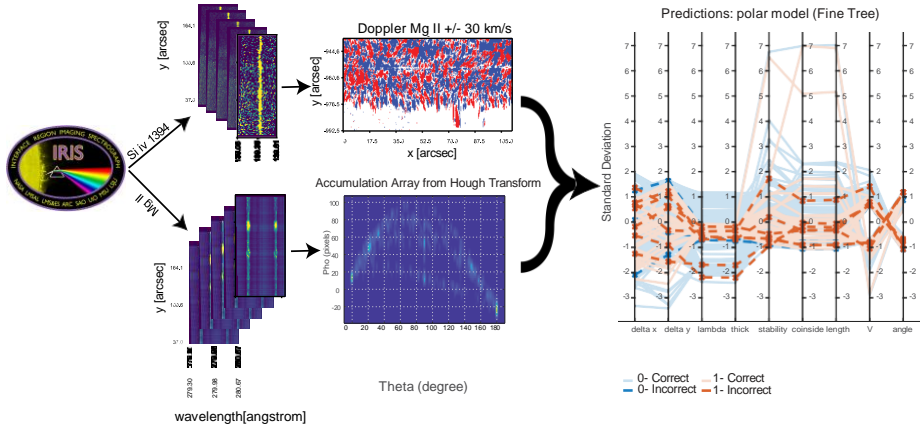

**Figure 1.** IRIS provided spectral data from 2013 onwards, enabling a detailed study of twisted spicules and their formation. The combination of Hinode and IRIS data, along with their spectra, deepened the understanding of these dynamic processes. Preprocessing steps were outlined to ensure the analysis-ready state of the IRIS dataset, facilitating the extraction and analysis of spicule features.

## 5. Discussion

The study on spicules and their role in solar wind energy and dynamics yielded significant findings:

The transition region (TR) of the Sun exhibits strong torsional structure, with distinct spicule populations at the poles and equator, challenging previous assumptions about helix jets dominating the TR. Spicule observations indicate a higher density in polar regions compared to the equatorial regions, with a slight asymmetry favoring the east side of the Sun over the west side. Rotating spicules are more prevalent in polar regions, suggesting their significant involvement in generating solar wind energy. The abundance of rotating spicules near the solar poles and in coronal hole areas is attributed to the rupture of magnetic mini-loops, which possess magnetic helicity and drive the rotational and helical motion of spicules. The origin and formation mechanisms of spicules remain incompletely understood, but mini-loop interchange reconnection is proposed as a potential process involved in spicule formation. Detailed and systematic observations of the low-coronal region, combined with the outer corona, are crucial for comprehending the origins of fast solar wind and mini-loops. Tornado-type eruptions, including surges, consist of fine, elongated threads and are associated with helical chromospheric jets, adding complexity to the transition from the corona to the fast solar wind. Twisting spicules are associated with high-speed polar winds and play a critical role in particle acceleration and mass loading in fast solar winds. Overall, this study provides valuable insights into the intricate dynamics of spicules and their contributions to solar wind energy. It highlights the importance of further research in this field to deepen our understanding.

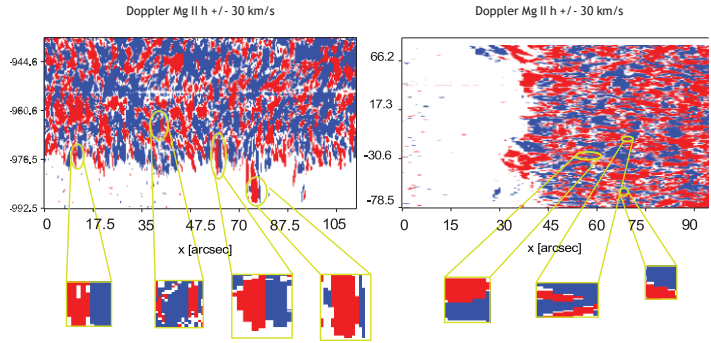

**Figure 2.** The labeling of twisted spicules on Doppler maps was highlighted, emphasizing their unique color pattern. Doppler maps are generated by examining spectral line shifts, revealing line-of-sight velocities. Twisted spicules exhibit distinct color patterns on the Doppler map, featuring adjacent red and blue areas without separation, indicating intertwined opposite velocity flows within the spicules.

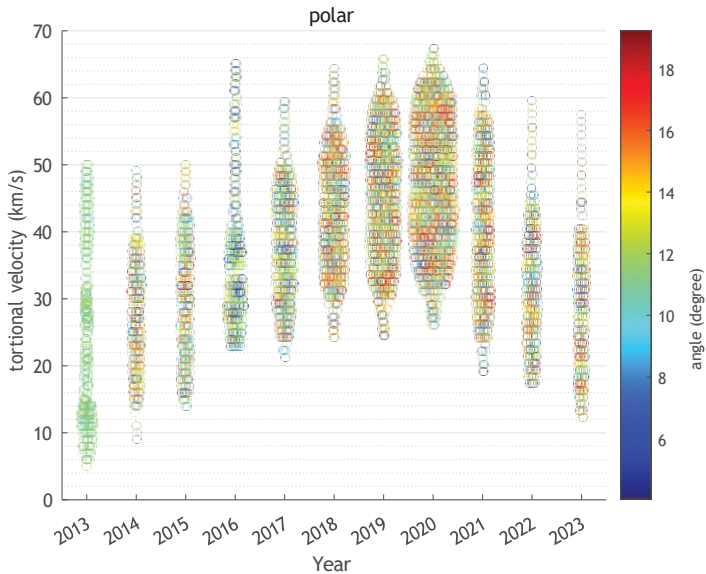

**Figure 3.** The significant difference in the number and velocity of rotational spicules between polar and equatorial regions has important implications for understanding their connection to material supply for fast solar winds and their acceleration mechanisms. Polar regions exhibit a higher occurrence of rotational spicules with greater velocities, suggesting they may serve as the primary source of fast solar winds. In contrast, equatorial regions have fewer rotational spicules with lower velocities, indicating distinct processes as the source of fast solar winds. Focusing on initial signals from polar activity can provide valuable insights into the activity originating from these regions.

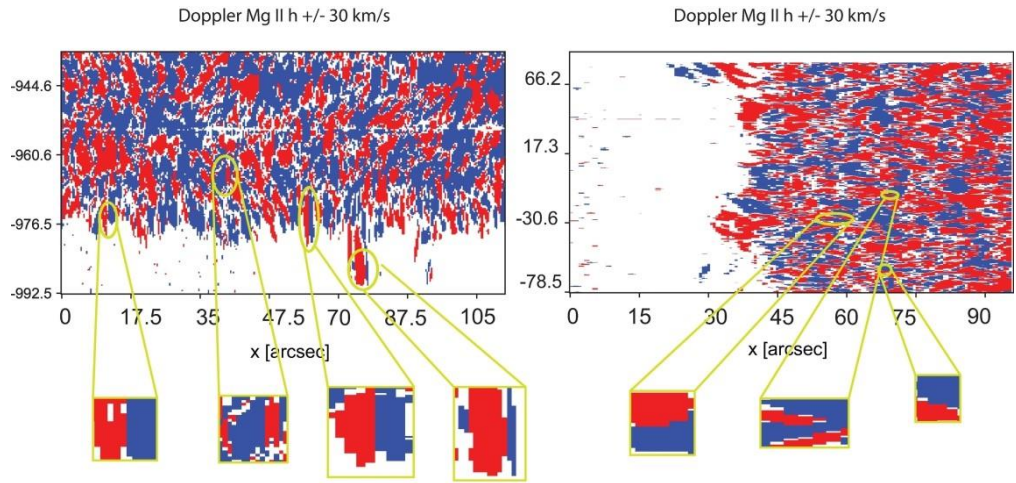

The labeling of twisted spicules on Doppler maps was highlighted, emphasizing their unique color pattern. Doppler maps are generated by examining spectral line shifts, revealing line-of-sight velocities. Twisted spicules exhibit distinct color patterns on the Doppler map, featuring adjacent red and blue areas without separation, indicating intertwined opposite velocity flows within the spicules.

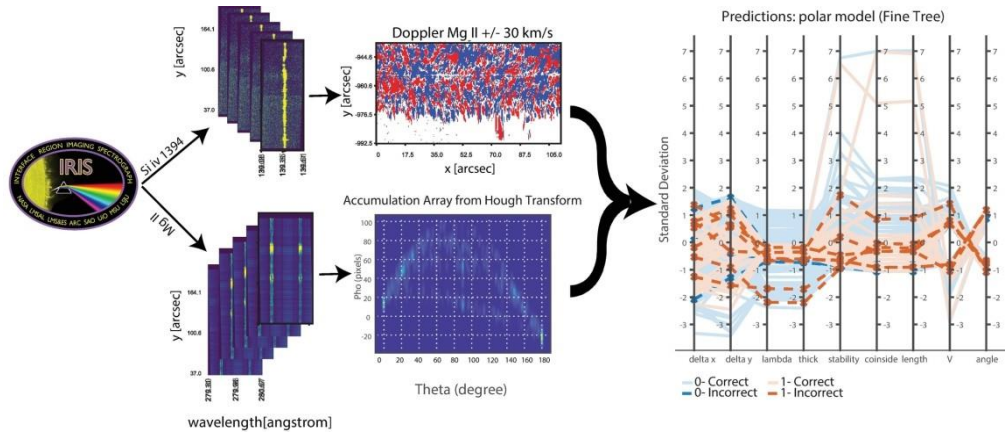

IRIS provided spectral data from 2013 onwards, enabling a detailed study of twisted spicules and their formation. The combination of Hinode and IRIS data, along with their spectra, deepened the understanding of these dynamic processes. Preprocessing steps were outlined to ensure the analysis-ready state of the IRIS dataset, facilitating the extraction and analysis of spicule features.

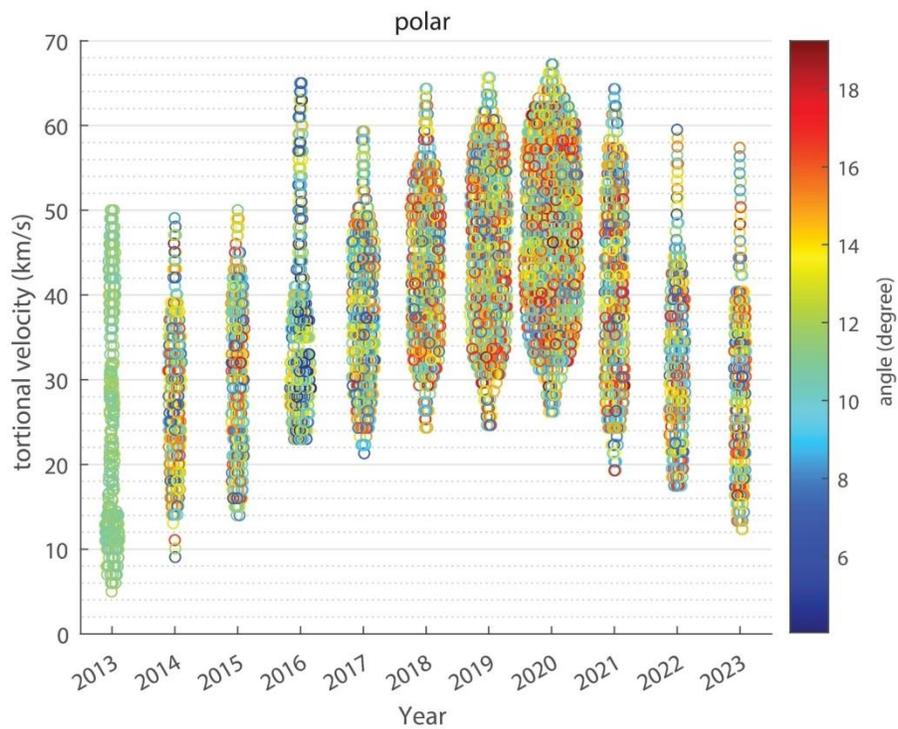

The significant difference in the number and velocity of rotational spicules between polar and equatorial regions has important implications for understanding their connection to material supply for fast solar winds and their acceleration mechanisms. Polar regions exhibit a higher occurrence of rotational spicules with greater velocities, suggesting they may serve as the primary source of fast solar winds. In contrast, equatorial regions have fewer rotational spicules with lower velocities, indicating distinct processes as the source of fast solar winds. Focusing on initial signals from polar activity can provide valuable insights into the activity originating from these regions.